\documentclass[12pt]{article}

\usepackage{amssymb,amsmath,url}

\numberwithin{equation}{section}

\newenvironment{Proof}{\removelastskip\par\medskip
\noindent{\em Proof.} \rm}{\penalty-20\null\hfill$\square$\par\medbreak}

\newenvironment{Proofy}{\removelastskip\par\medskip
\noindent{\em Proof} \hskip-0.1cm \rm}{\penalty-20\null\hfill$\square$\par\medbreak}

\allowdisplaybreaks

 \def\real{{\mathord{\mathbb R}}}
 
 \def\inte{{\mathord{\mathbb N}}}
 
 \def\qu{{\mathord{\mathbb Z}}}

 \def\Dom{{\mathrm{{\rm Dom}}}}

 \def\real{{\mathord{{\rm I\kern-3pt R}}}}        % Fake blackboard bold R.
 
 \def\inte{{\mathord{{\rm I\kern-3pt N}}}}
 \def\sZZ{{\rm Z\kern-.45em{}Z}}

 \def\sQQ{{\kern 0.27em \vrule height1.45ex width0.03em depth0em
           \kern-0.30em \rm Q}}
 \def\qu{{\mathchoice
         {\sQQ}
         {\sQQ}
   {\kern 0.225em \vrule height1.05ex width0.025em depth0em \kern-0.25em \rm Q}
   {\kern 0.180em \vrule height0.78ex width0.020em depth0em \kern-0.20em \rm Q}
         }}
 \def\sGG{{\kern 0.27em \vrule height1.45ex width0.03em depth0em
           \kern-0.30em \rm G}}
 \def\gg{{\mathchoice
         {\sGG}
         {\sGG}
   {\kern 0.225em \vrule height1.05ex width0.025em depth0em \kern-0.25em \rm G}
   {\kern 0.180em \vrule height0.78ex width0.020em depth0em \kern-0.20em \rm G}
         }}

 \newtheorem{prop}{Proposition}[section]
 \newtheorem{lemma}[prop]{Lemma}
 
 \newtheorem{corollary}[prop]{Corollary}

 \def\Dom{{\mathrm{{\rm Dom \! \ }}}}
 \def\P{{\mathord{\mathbb P}}}

\def\E{\mathop{\hbox{\rm I\kern-0.20em E}}\nolimits}

 \newcounter{hyp}
\title{\huge 
 Hedging in bond markets by the Clark-Ocone formula 
} 
 
\author{ 
Nicolas Privault\thanks{nprivault@ntu.edu.sg}
\\ 
\vspace{-0.1cm} 
\small 
School of Physical and Mathematical Sciences\\
\vspace{-0.1cm} 
\small 
Nanyang Technological University\\ 
\vspace{-0.1cm} 
\small 
SPMS-MAS, 21 Nanyang Link\\ 
\vspace{-0.1cm} 
\small 
Singapore 637371 
\and 
Timothy Robin Teng 
\thanks{timoteng@mathsci.math.admu.edu.ph} 
\\
\vspace{-0.1cm} 
\small 
Department of Mathematics\\
\vspace{-0.1cm} 
\small 
Ateneo de Manila University\\ 
\vspace{-0.1cm} 
\small 
Loyola Heights \\ 
\vspace{-0.1cm} 
\small 
Quezon City, Philippines 
}

\begin{document}

\hyphenation{func-tio-nals} 
\hyphenation{Privault} 

\maketitle 

\vspace{-1cm} 
 
\begin{abstract}
{ 
Hedging strategies in bond markets are computed by martingale representation and the Clark-Ocone formula under the choice of a suitable of numeraire, in a model driven by the dynamics of bond prices. Applications are given to the hedging of swaptions and other interest rate derivatives, and our approach is compared to delta hedging when the underlying swap rate is modeled by a diffusion process. 
} 
\end{abstract} 
 
\baselineskip0.7cm
 
\section{Introduction} 
 The pricing of interest rate derivatives is usually 
 performed by the change of numeraire technique under 
 a suitable forward measure $\hat{\P}$. 
 On the other hand, 
 the computation of hedging strategies for interest rate derivatives 
 presents several difficulties, 
 in particular, hedging strategies appear not to be unique 
 and one is faced with the problem of choosing an appropriate 
 tenor structure of bond maturities 
 in order to correctly hedge maturity-related risks, 
 see e.g. \cite{corcuera} in the jump case. 
\\ 
 
 In this paper we consider the application of the change of 
 numeraire technique to the computation of hedging strategies 
 for interest rate derivatives. 
 The payoff of an interest derivative is usually based 
 on an underlying asset priced $\hat{X}_t$ at time $t$ 
 (e.g. a swap rate) which is 
 defined from a family $(P_t(T_i))_i$ of bond prices 
 with maturities $(T_i)_i$. 
\\ 
 
 In this paper we distinguish between two different modeling situations. 
\begin{description} 
\item{(1)} Modeling $\hat{X}_t$ as a Markov diffusion process 
\begin{equation} 
\label{dx1} 
 d\hat{X}_t = \hat{\sigma}_t ( \hat{X}_t ) 
 d \hat{W}_t 
\end{equation} 
 where $(\hat{W}_t )_{t\in \real_+}$ is a 
 Brownian motion under the forward measure $\hat{\P}$. 
 In this case delta hedging can be applied and 
 this approach has been adopted in \cite{jamshidian2} to compute 
 self-financing hedging strategies for swaptions based on 
 geometric Brownian motion. 
 In Section~\ref{mc} of this paper we review and extend this approach.  
\item{(2)} Modeling each bond price $P_t(T)$ by a stochastic differential 
 equation of the form 
\begin{equation}
\label{djld1} 
 dP_t (T) 
 = 
 r_t P_t (T) dt 
 + 
 P_t (T) \zeta_t (T) dW_t, 
\end{equation} 
 where $W_t$ is a standard Brownian motion under the risk-neutral 
 measure $\P$. 
 In this case the process $\hat{X}_t$ 
 may no longer have a simple Markovian dynamics 
 under $\hat{\P}$ 
 (cf. Lemma~\ref{l} or \eqref{wsht} below) and 
 we rely on the Clark-Ocone formula which is 
 commonly used for the hedging of path-dependent options. 
 Precisely, due to the use of forward measures 
 we will apply the Clark-Ocone formula under 
 change of measure of \cite{ko:gc}. 
 This approach is carried out in Section~\ref{6}. 
\end{description} 
 We consider a bond price curve $(P_t)_{t\in \real_+}$, 
 valued in a real separable Hilbert space $G$, usually a weighted 
 Sobolev space of real-valued functions on $\real_+$, 
 cf. \cite{filipovicbk} and \S~6.5.2 of \cite{carmonatehranchi}, 
 and we denote by 
 $G^*$ the dual space of continuous linear mappings on $G$. 
\\ 
 
 Given $\mu \in G^*$ a signed finite measure 
 on $\real_+$ with support in $[T,\infty )$, 
 we consider 
$$ 
 P_t(\mu) 
 : = 
 \langle \mu , P_t \rangle_{G^* \! ,G} 
 = 
 \int_T^\infty 
 P_t(y) \mu (d{y}) 
, 
$$ 
 which represents a basket of bonds whose maturities are 
 beyond the exercise date $T>0$ and distributed according to the measure $\mu$. 
 The value of a portfolio strategy 
 $(\phi_t )_{t\in [0,T]}$ is given by 
\begin{equation} 
\label{000} 
 V_t : = 
 \langle 
 \phi_t , 
 P_t 
 \rangle_{G^* \! ,G} 
 = 
 \int_T^\infty 
 P_t(y) \phi_t (d{y}) 
\end{equation} 
 where the measure $\phi_t (d{y})$ represents the amount of 
 bonds with maturity in $[{y},{y}+d{y}]$ in the portfolio 
 at time $t \in [ 0 , T]$. 
\\ 
 
 Given $\nu \in G^*$ another positive finite measure on $\real_+$ 
 with support in $[T,\infty )$, we consider the generalized 
 annuity numeraire 
$$ 
 P_t ( \nu ) : = 
 \langle \nu , P_t \rangle_{G^* \! ,G} 
 = 
 \int_T^\infty 
 P_t(y) \nu (d{y}) 
, 
$$ 
 and the forward bond price curve 
$$ 
 \hat{P}_t 
 = 
 \frac{P_t}{P_t(\nu)}  
, 
 \qquad 
 0\leq t \leq T, 
$$ 
 which is a martingale under 
 the forward measure $\hat{\P}$ defined by 
\begin{equation} 
\label{fm} 
 \E \left[ 
 \frac{d\hat{\P}}{d\P } 
 \Big| 
 {\cal F}_S 
 \right] 
 = 
 e^{-\int_0^S r_s ds } 
 \frac{P_S(\nu)}{P_0(\nu)}, 
\end{equation}  
 where the maturity $S$ is such that $S \geq T$. 
\\ 
 
 In practice, $\mu (dy)$ and $\nu (dy)$ 
 will be finite point measures, i.e. sums 
$$ 
 \sum_{k=i}^j \alpha_k \delta_{T_k} ( dy ) 
$$ 
 of Dirac measures based on the maturities $T_i, \ldots , T_j \geq T$ 
 of a given a tenor structure, in which $\alpha_k$ represents 
 the amount allocated to a bond with maturity $T_k$, 
 $k = i , \ldots , j$. 
 In this case we are interested in finding a hedging 
 strategy $\phi_t (dy)$ of the form 
$$\phi_t ( dy ) = 
 \sum_{k=i}^j 
 \alpha_k ( t ) 
 \delta_{T_k} (dy) 
$$ 
 in which case \eqref{000} reads 
$$ 
 V_t = 
 \sum_{k=i}^j 
 \alpha_k ( t ) 
 P_t (T_k) 
, \qquad 
 0 \leq t \leq T, 
$$ 
 and similarly for $P_t(\mu)$ and $P_t(\nu)$ 
 using $\mu (dx)$ and $\nu (dx)$ 
 respectively. 
\\ 
 
 Lemma~\ref{p02} below shows how to compute self-financing 
 hedging strategies from the decomposition 
\begin{equation} 
\label{djklfff} 
 \hat{\xi} = 
 \hat{\E} [ \hat{\xi} ] 
 + 
 \int_0^T \langle {}\phi_s , d\hat{P}_s \rangle_{G^* \! ,G} 
, 
\end{equation} 
 of a forward claim payoff $\hat{\xi} = \xi / P_S (\nu)$, where 
 $({}\phi_t)_{t\in [0,T]}$ is a square-integrable 
 $G^*$-valued adapted process of continuous linear mappings on $G$. 
 The representation \eqref{djklfff} can be obtained from 
 the predictable representation 
\begin{equation} 
\label{fkllff}
 \hat{\xi} 
 = \hat{\E } [\hat{\xi} ] 
 + \int_0^T 
 \langle \hat{\alpha}_t , d\hat{W}_t \rangle_H 
, 
\end{equation} 
 where $(\hat{W}_t )_{t\in \real_+}$ is a 
 Brownian motion under $\hat{\P}$ with 
 values in a separable Hilbert space $H$, cf. 
 \eqref{*2} below, and $(\hat{\alpha}_t)_{t\in \real_+}$ 
 is an $H$-valued square-integrable ${\cal F}_t$-adapted process. 
\\ 
 
 In case the forward price process 
 $\hat{P}_t = {P_t}/{P_t(\nu)}$, $t\in \real_+$, 
 follows the dynamics 
\begin{equation} 
\label{isc} 
 d \hat{P}_t = \hat{\sigma}_t d\hat{W}_t 
, 
\end{equation} 
 where $(\hat{\sigma}_t)_{t\in \real_+}$ 
 is an ${\cal L}_{HS} ( H , G)$-valued adapted process of 
 Hilbert-Schmidt operators from $H$ to $G$, 
 cf. 
 \cite{carmonatehranchi}, 
 and $\hat{\sigma}_t^* : H \to G^*$ 
 is invertible, $0 \leq t \leq T$, 
 Relation~\eqref{isc} shows that the process 
 $({}\phi_t)_{t\in \real_+}$ in Lemma~\ref{p02} is given by 
\begin{equation} 
\label{asfghfsd} 
 {}\phi_t 
 = 
 ( \hat{\sigma}_t^*)^{-1} 
 \hat{\alpha}_t, 
 \qquad 
 0 \leq t \leq T. 
\end{equation} 
 However this invertibility condition can be too restrictive in 
 practice. 
\\ 
 
 On the other hand the invertibility of $\sigma^*_t : G^* \to H$ as 
 an operator is 
 not required in order to hedge the claim $\xi$. 
 As an illustrative example, when $H=\real$ 
 we have 
$$ 
 \hat{\xi} 
 = \E [ \hat{\xi} ] 
 + \int_0^T 
 \hat{\alpha}_t d\hat{W}_t 
 = 
 \E [ \hat{\xi} ] 
 + 
 \sum_{i=1}^n 
 c_i 
 \int_0^T 
 \frac{\hat{\alpha}_t}{\hat{\sigma}_t (T_i)} 
 d\hat{P}_t (T_i) 
, 
$$ 
 where $\{T_1,\ldots ,T_n\} \subset \real_+$ is a given 
 tenor structure and 
 $c_1,\ldots ,c_n \in \real_+$ satisfy $c_1+\cdots + c_n=1$, 
 and we can take 
$$ 
 {}\phi_t = 
 \sum_{i=1}^n 
 c_i 
 \frac{\hat{\alpha}_t}{\hat{\sigma}_t (T_i)} 
 \delta_{T_i} 
. 
$$ 
 Such a hedging strategy $(\phi_t)_{t\in [0,T]}$ depends 
 as much on the bond structure 
 (through the volatility process $\sigma_t ( x )$) 
 as on the claim $\xi$ itself (through $\alpha_t$), in connection 
 with 
 the problem of hedging maturity-related risks. 
\\ 
 
 The predictable representation \eqref{fkllff} can 
 be computed from the Clark-Ocone formula for the 
 Malliavin gradient $\hat{D}$ with respect to 
 $(\hat{W}_t)_{t\in \real_+}$, 
 cf. e.g. Proposition~6.7 in \S~6.5.5 of \cite{carmonatehranchi} 
 when the numeraire is the money market account, 
 cf. also \cite{privault-teng2} for examples of explicit 
 calculations in this case. 
 This approach is more suitable to a non-Markovian or path-dependent 
 dynamics specified for $(\hat{P}_t)_{t\in \real_+}$ as a functional 
 of $(\hat{W}_t)_{t\in \real_+}$. 
 However this is not the approach chosen here 
 since the dynamics assumed for the bond price is either 
 Markovian as in \eqref{dx1}, cf. Section~\ref{mc}, 
 or written in terms of $W_t$ as in \eqref{djld1}, 
 cf. Section~\ref{6}. 
\\ 
 
 In this paper we specify the dynamics of $(P_t)_{t\in \real_+}$ 
 under the risk-neutral measure and we apply the 
 Clark-Ocone formula under a change of measure \cite{ko:gc}, 
 using the Malliavin gradient $D$ with respect to 
 $W_t$, cf. \eqref{adfdsg} below. 
 In Proposition~\ref{p52} below we compute self-financing 
 hedging strategies for contingent claims with 
 payoff of the form $ \xi 
 = 
 P_S (\nu) \hat{g} 
 \left( 
 {P_T(\mu)} / {P_T(\nu)} 
 \right)$. 
\\ 
 
 This paper is organized as follows. 
 Section~\ref{2} 
 contains the preliminaries on 
 the derivation of 
 self-financing hedging strategies 
 by change of numeraire 
 and the Clark-Ocone formula under change of measure. 
 In Section~\ref{6} we use the Clark-Ocone formula under a 
 change of measure to compute self-financing hedging strategies for 
 swaptions and other derivatives based on the dynamics of $(P_t)_{t\in \real_+}$. 
 In Section~\ref{mc} we compare the above results with 
 the delta hedging approach 
 when the dynamics of the swap rate $(\hat{X}_t)_{t\in \real_+}$ 
 is based on a diffusion process. 
\section{Preliminaries} 
\label{ss2} 
 In this section we review the hedging of options 
 by change of numeraire, cf. e.g. \cite{geman}, \cite{protterspa}, 
 in the 
 framework of 
 \cite{carmonatehranchi}. 
 We also quote the Clark-Ocone formula under change of measure. 
\subsubsection*{
 Hedging by change of numeraire} 
\label{2} 
 Consider a numeraire $(M_t)_{t\in \real_+}$ under the 
 risk-neutral probability measure $\P$ on 
 a filtered probability space $(\Omega , ({\cal F}_t)_{t\in\real_+} , \P )$, 
 that is, $(M_t)_{t\in \real_+}$ is a continuous, strictly positive, 
 ${\cal F}_t$-adapted asset price process 
 such that the discounted price process 
 $e^{-\int_0^t r_s ds} M_t$ is an ${\cal F}_t$-martingale 
 under $\P$. 
\\ 
 
 Recall that an option with payoff $\xi$, 
 exercise date $T$ and maturity $S$, 
 is priced at time $t$ as 
\begin{equation} 
\label{*1} 
 \E\left[ 
 e^{-\int_t^S r_s ds} 
 \xi \ 
 \Big| 
 {\cal F}_t 
 \right] 
 = 
 M_t 
 \hat{\E}
 [ 
 \hat{\xi} 
 | 
 {\cal F}_t 
 ] 
, 
 \qquad 
 0 \leq t \leq T, 
\end{equation} 
 under the forward measure 
 $\hat{\P}$ defined by 
\begin{equation} 
\label{fwd} 
 \E \left[ 
 \frac{d\hat{\P}}{d\P } 
 \Big| 
 {\cal F}_S 
 \right] 
 = 
 e^{-\int_0^S r_s ds } 
 \frac{M_S}{M_0}, 
\end{equation} 
 $S\geq T$, where 
$$ 
 \hat{\xi} = \frac{\xi}{M_S} 
 \in L^1 ( \hat{\P} , {\cal F}_S ) 
$$ 
 denotes the forward payoff of the claim $\xi$. 
\\ 
 
 In the 
 framework of 
 \cite{carmonatehranchi}, consider 
 $(W_t)_{t\in \real_+}$ a cylindrical Brownian motion 
 taking values in a separable Hilbert space $H$ with covariance 
$$ 
 E [ W_s ( h ) W_t ( k ) ] = ( s \wedge t ) \langle h , k \rangle_H, 
 \qquad 
 h , k \in H, 
 \quad 
 s, t \in \real_+ 
, 
$$ 
 and generating the filtration $({\cal F}_t)_{t\in \real_+}$. 
 Consider a continuous ${\cal F}_t$-adapted 
 asset price process $(X_t)_{t\in \real_+}$ 
 taking values in a real separable Hilbert 
 space $G$, and assume that 
 both $(X_t)_{t\in \real_+}$ and $(M_t)_{t\in \real_+}$ 
 are It\^o processes in the sense of \S~4.2.1 
 of \cite{carmonatehranchi}. 
 The forward asset price 
$$ 
 \hat{X}_t : = \frac{X_t}{M_t}, \qquad 0 \leq t \leq T, 
$$ 
 is a martingale in $G$ under the forward measure $\hat{\P}$, 
 provided it is integrable under $\hat{\P}$. 
\\ 
 
 The next lemma will be key to compute 
 self-financing portfolio strategies in the assets 
 $(X_t,M_t)$ by numeraire invariance, cf. \cite{protterspa}, \cite{cfhuang} 
 for the finite dimensional case. 
 We say that a portfolio $({}\phi_t, {}\eta_t)_{t\in [0,T]}$ 
 with value 
$$ 
 \langle 
 {}\phi_t 
 , 
 X_t 
 \rangle_{G^* \! ,G} 
 + 
 {}\eta_t  
 M_t 
, 
 \qquad 
 0 \leq t \leq T, 
$$ 
 is self-financing if 
\begin{equation} 
\label{isf} 
 dV_t 
 = 
 \langle {}\phi_t , d X_t \rangle_{G^* \! ,G} 
 + 
 {}\eta_t dM_t 
. 
\end{equation} 
 The portfolio $({}\phi_t, {}\eta_t)_{t\in [0,T]}$ 
 is said to hedge the claim $\xi = M_S \hat{\xi}$ if 
$$ 
 \langle 
 {}\phi_t 
 , 
 X_t 
 \rangle_{G^* \! ,G} 
 + 
 {}\eta_t  
 M_t 
 = 
 \E\left[ 
 e^{-\int_t^S r_s ds} 
 M_S 
 \hat{\xi} 
 \ \Big| 
 {\cal F}_t 
 \right] 
, 
 \qquad
 0 \leq t \leq T. 
$$ 
\begin{lemma} 
\label{p02} 
 Assume that the forward claim price 
$ 
 \hat{V}_t 
 : =  \hat{\E} 
 [ 
 \hat{\xi} 
 | 
 {\cal F}_t 
 ] 
$ 
 has the predictable representation 
\begin{equation} 
\label{prd} 
 \hat{V}_t = 
 \hat{\E} [ \hat{\xi} ] 
 + 
 \int_0^t \langle {}\phi_s , d\hat{X}_s \rangle_{G^* \! ,G} 
, \qquad 
 0 \leq t \leq T, 
\end{equation} 
 where 
 $({}\phi_t)_{t\in [0,T]}$ is a square-integrable 
 $G^*$-valued adapted process of continuous linear mappings on $G$. 
 Then the portfolio $({}\phi_t, {}\eta_t)_{t\in [0,T]}$ 
 defined with 
\begin{equation} 
\label{dw} 
 {}\eta_t = \hat{V}_t - \langle {}\phi_t , \hat{X}_t \rangle_{G^*,G}, 
 \qquad 
 0 \leq t \leq T 
, 
\end{equation} 
 and priced as 
$$ 
 V_t 
 = 
 \langle 
 {}\phi_t 
 , 
 X_t 
 \rangle_{G^* \! ,G} 
 + 
 {}\eta_t  
 M_t 
, 
 \qquad 
 0 \leq t \leq T, 
$$ 
 is self-financing and hedges the claim $\xi = M_S \hat{\xi}$. 
\end{lemma} 
\begin{Proof} 
 For completeness we provide the proof of this lemma, 
 although it is a direct 
 extension of classical results. 
 In order to check that the portfolio 
 $({}\phi_t, {}\eta_t)_{t\in [0,T]}$ 
 hedges the claim $\xi = M_S \hat{\xi}$ it suffices to note that 
 by \eqref{*1} and \eqref{dw} we have 
$$ 
 \langle 
 {}\phi_t 
 , 
 X_t 
 \rangle_{G^* \! ,G} 
 + 
 {}\eta_t  
 M_t 
 = 
 M_t 
 \hat{V}_t 
 = 
 \E\left[ 
 e^{-\int_t^S r_s ds} 
 M_S 
 \hat{\xi} 
 \ \Big| 
 {\cal F}_t 
 \right] 
, 
 \qquad
 0 \leq t \leq T. 
$$ 
 The portfolio $({}\phi_t, {}\eta_t)_{t\in [0,T]}$ 
 is clearly self-financing for $(\hat{X}_t,1)$ by \eqref{prd}, 
 and by the semimartingale version of numeraire invariance, 
 cf. e.g. page 184 of \cite{protterspa}, and \cite{cfhuang}, 
 it is also self-financing for $(X_t,M_t)$. 
\\ 
 
 cf. also \S~3.2 of \cite{jamshidian4} and references therein. 
\\ 
 
 For completeness we quote the proof of the self-financing property, 
 as follows: 
\begin{eqnarray} 
\nonumber 
\lefteqn{ 
 dV_t 
 = 
 d ( M_t \hat{V}_t ) 
} 
\\ 
\nonumber 
 & = & 
 \hat{V}_t dM_t 
 + 
 M_t d \hat{V}_t 
 + 
 d M_t \cdot d \hat{V}_t 
\\ 
\nonumber 
 & = & 
 \hat{V}_t dM_t 
 + 
 M_t \langle {}\phi_t , d\hat{X}_t \rangle_{G^*,G} 
 + 
 d M_t \cdot \langle {}\phi_t , d\hat{X}_t \rangle_{G^*,G} 
\\ 
\nonumber 
 & = & 
 \langle {}\phi_t , \hat{X}_t \rangle_{G^*,G} 
 dM_t 
 + 
 M_t 
 \langle {}\phi_t , d\hat{X}_t \rangle_{G^*,G} 
 + 
 dM_t \cdot \langle {}\phi_t , d\hat{X}_t \rangle_{G^*,G} 
 + 
 ( \hat{V}_t - 
 \langle {}\phi_t , \hat{X}_t \rangle_{G^*,G} 
 ) dM_t 
\\ 
\nonumber 
 & = & 
 \langle {}\phi_t , d ( M_t \hat{X}_t ) \rangle_{G^*,G} 
 + 
 ( \hat{V}_t - 
 \langle {}\phi_t , \hat{X}_t \rangle_{G^*,G} 
 ) dM_t 
\\ 
\nonumber 
 & = & 
 \langle {}\phi_t , d X_t \rangle_{G^* \! ,G} 
 + 
 {}\eta_t dM_t 
. 
\end{eqnarray} 
\end{Proof} 
 Lemma~\ref{p02} yields a self-financing 
 portfolio $( {}\phi_t , {}\eta_t )_{t \in [0,T]}$ with value 
\begin{equation} 
\label{jklf} 
 V_t 
 = 
 V_0 
 + 
 \int_0^t {}\eta_s dM_s + \int_0^t \langle {}\phi_s , d X_s \rangle_{G^* \! ,G} 
, 
 \qquad 
 0 \leq t \leq T, 
\end{equation} 
 given by \eqref{isf}, 
 which hedges the claim with exercise date $T$ and random payoff $\xi$. 
\subsubsection*{Clark formula under change of measure} 
 Recall that by the Girsanov theorem, cf. Theorem~10.14 of \cite{daprato} 
 or Theorem~4.2 of \cite{carmonatehranchi}, 
 the process $(\hat{W}_t)_{t\in \real_+}$ defined by 
\begin{equation} 
\label{*2} 
 d \hat{W}_t 
 = 
 d W_t 
 - 
 \frac{1}{M_t} 
 d M_t \cdot dW_t 
, 
 \qquad 
 t \in \real_+ 
, 
\end{equation} 
 is a $H$-valued Brownian motion under $\hat{\P}$. 
 Let $D$ denote the Malliavin gradient with respect to 
 $(W_t)_{t\in \real_+}$, defined on smooth functionals 
$$ 
 \hat{\xi} = f( W_{t_1},\ldots , W_{t_n}) 
$$ 
 of Brownian motion, $f \in {\cal C}_b ( \real^n)$, as 
$$ 
 D_t \hat{\xi} = \sum_{k=1}^n 
 {\bf 1}_{[0,t_k]} (t ) 
 \frac{\partial f}{\partial x_k} 
 (W_{t_1},\ldots , W_{t_n}), 
 \qquad 
 t\in \real_+, 
$$ 
 and extended by closability to its domain $\Dom (D)$. 
 The proof of Proposition~\ref{p52} 
 relies on the following Clark-Ocone formula under a change of 
 measure, cf. 
 \cite{ko:gc}, 
 which can be extended to $H$-valued Brownian motion by 
 standard arguments. 
\begin{lemma} 
\label{cf} 
 Let $(\gamma_t)_{t\in \real_+}$ denote a $H$-valued 
 square-integrable ${\cal F}_t$-adapted process 
 such that $\gamma_t \in \Dom (D)$, $t\in \real_+$, and 
\begin{equation*} 
 dW_t = \gamma_t dt + d\hat{W}_t
. 
\end{equation*} 
 Let $\hat{\xi} \in \Dom (D)$ such that 
\begin{equation} 
\label{2.k.0} 
 \hat{E} \left[ \int_0^T \Vert D_t \hat{\xi} \Vert_H^2 dt \right] < \infty 
\end{equation} 
 and 
\begin{equation} 
\label{1.k.0} 
 \hat{E} \left[ 
 | \hat{\xi} | 
 \int_0^T 
 \left\| 
 \int_0^T 
 D_t \gamma_s d\hat{W}_s 
 \right\|^2_H 
 dt 
 \right] < \infty 
. 
\end{equation} 
 Then the predictable representation 
$$ 
 \hat{\xi} = \hat{\E} [\hat{\xi} ] + \int_0^T 
 \langle \hat{\alpha}_t , d\hat{W}_t \rangle_H 
$$ 
 is given by 
\begin{equation} 
\label{adfdsg} 
 \hat{\alpha}_t 
 = 
 \hat{\E} 
 \left[ 
 D_t\hat{\xi} 
 + 
 \hat{\xi} \int_t^T D_t \gamma_s
 d\hat{W}_s 
 \Big| 
 {\cal F}_t 
 \right] 
, 
 \qquad 
 0 \leq t \leq T. 
\end{equation} 
\end{lemma}
\section{Hedging by the Clark-Ocone formula} 
\label{6} 
 In this section we present a 
 computation of hedging strategies using the Clark-Ocone formula 
 under change of measure and 
 we assume that the dynamics of $(P_t)_{t\in \real_+}$ is given by 
 the 
 stochastic differential equation 
\begin{equation} 
\label{cfp12} 
 dP_t 
 = 
 r_t P_t dt 
 + 
 P_t \zeta_t dW_t 
, 
\end{equation} 
 in the Sobolev space $G$ which is assumed to be 
 an algebra of real-valued functions on $\real_+$, 
 and $(\zeta_t)_{t\in \real_+}$ is 
 an ${\cal L}_{HS} ( H , G )$-valued deterministic function. 
\\ 
 
 The aim of this section is to prove Proposition~\ref{p52} 
 below under the non-restrictive integrability conditions 
\begin{equation} 
\label{1.k} 
 \int_0^T 
 \int_T^\infty 
 \Vert \zeta_t ( {y} ) \Vert^2_H 
 \hat{\E} [ | \hat{P}_T |^2 (y) ] 
 \mu ( d{y} ) 
 dt 
 < \infty 
\end{equation}  
 and 
\begin{equation} 
\label{2.k} 
 \int_0^T 
 \int_T^\infty 
 \Vert \zeta_t ( {y} ) \Vert^2_H 
 \hat{\E} 
 [ 
 | \hat{P}_T(\mu ) |^2 
 ( 
 |\hat{P}_T|^2 (y) 
 + 
 |\hat{P}_t|^2 (y) 
 )
 ] 
 \nu (d{y}) 
 dt 
 < \infty . 
\end{equation} 
 which are respectively derived from 
 \eqref{2.k.0} and \eqref{1.k.0}. 
 The next proposition provides an alternative to 
 Proposition~3.3 in \cite{privault-teng2} by applying 
 to a different family of payoff functions. 
 It coincides with Proposition~3.3 of 
 \cite{privault-teng2} in case $S=T$ and $\nu = \delta_T$. 
\begin{prop} 
\label{p52} 
 Consider the claim with payoff 
$$ 
 \xi 
 = 
 P_S (\nu) \hat{g} 
 \left( 
 \frac{P_T(\mu)}{P_T(\nu)} 
 \right), 
$$ 
 where $\hat{g} : \real \to \real$ 
 is a Lipschitz function. 
 Then the portfolio 
\begin{equation} 
\label{hk} 
 \phi_t ( d{y} ) 
 = 
 \hat{\E} \left[ 
 \frac{\hat{P}_T({y}) }{\hat{P}_t( y )  } 
 \hat{g}' ( \hat{P}_T(\mu ) )  
 \Big| 
 {\cal F}_t \right] 
 \mu ( d{y} ) 
 + 
 \hat{\E} \left[ 
 ( 
 \hat{g} ( \hat{P}_T(\mu ) )  
 - 
 \hat{P}_T(\mu ) 
 \hat{g}' ( \hat{P}_T(\mu ) )  
 ) 
 \frac{\hat{P}_T({y}) }{\hat{P}_t( y ) } 
 \Big| 
 {\cal F}_t \right] 
 \nu ( d {y} ) 
\end{equation} 
 $0\leq t \leq T$, 
 is self-financing and hedges the claim $\xi$. 
\end{prop} 
 Before proving Proposition~\ref{p52} we check that 
 the portfolio ${}\phi_t$ hedges the claim 
 $\xi = P_S(\nu) \hat{g} ( \hat{P}_T(\mu ) )$ 
 by construction, since we have 
\begin{eqnarray} 
\nonumber 
 {V}_t - \langle {}\phi_t , {P}_t \rangle_{G^*,G} 
 & = & 
 P_t(\nu) 
 \hat{\E} 
 \left[ 
 \hat{g} 
 ( 
 \hat{P}_T(\mu ) 
 ) 
 \Big| 
 {\cal F}_t 
 \right] 
 - 
 \int_{T}^\infty P_t(y) \phi_t ( dy ) 
\\ 
\nonumber 
 & = & 
 P_t(\nu) 
 \hat{\E} \left[ 
 \hat{g} ( \hat{P}_T (\mu ) )  
 \Big| 
 {\cal F}_t \right] 
\\ 
\nonumber 
 & & 
 - 
 \int_{T}^\infty 
 \hat{\E} \left[ 
 \frac{\hat{P}_T({y}) }{\hat{P}_t( y )  } 
 \hat{g}' ( \hat{P}_T (\mu ) )  
 \Big| 
 {\cal F}_t \right] 
 P_t(y) 
 \mu ( d{y} ) 
\\ 
\nonumber 
 & & 
 - 
 \int_{T}^\infty 
 \hat{\E} \left[ 
 ( 
 \hat{g} (  \hat{P}_T(\mu ) )  
 - 
 \hat{P}_T(\mu ) 
 \hat{g}' ( \hat{P}_T(\mu ) )  
 ) 
 \frac{\hat{P}_T({y}) }{\hat{P}_t( y ) } 
 \Big| 
 {\cal F}_t \right] 
 P_t(y) \nu ( d {y} ) 
\\ 
\nonumber 
 & = & 
 - 
 P_t(\nu) 
 \int_{T}^\infty 
 \hat{\E} \left[ 
 \hat{P}_T({y}) 
 \hat{g}' ( \hat{P}_T (\mu) )  
 \Big| 
 {\cal F}_t \right] 
 \mu ( d{y} ) 
\\ 
\nonumber 
 & & 
 + 
 P_t(\nu) 
 \int_{T}^\infty 
 \hat{\E} \left[ 
 \hat{P}_T(\mu ) 
 \hat{g}' ( \hat{P}_T(\mu ) )  
 \hat{P}_T({y}) 
 \Big| 
 {\cal F}_t \right] 
 \nu ( d {y} ) 
\\ 
\label{bc} 
 & = & 
 0 
. 
\end{eqnarray} 
 The identity \eqref{bc} will also be used in the proof of 
 Lemma~\ref{l2b} below. 
\\ 
 
 Before moving to the proof of Proposition~\ref{p52} we 
 consider some examples of applications of the results of 
 Proposition~\ref{p52}, 
 in which the dynamics of $(P_t)_{t\in \real_+}$ is given by 
 \eqref{djld1}. 
\subsubsection*{Exchange options} 
 In the case of an exchange option 
 with $S=T$ and payoff 
 $(P_T(\mu) - \kappa P_T(\nu))^+$, 
 Proposition~\ref{p52} yields the self-financing 
 hedging strategy 
\begin{eqnarray*} 
 \phi_t ( d{y} ) 
 & = & 
 \hat{\E} \left[ 
 {\bf 1}_{\{ \hat{P}_T (\mu) > \kappa \} } 
 \frac{\hat{P}_T({y}) }{\hat{P}_t(y)  } 
 \Big| 
 {\cal F}_t \right] 
 \mu ( d{y} ) 
 - \kappa 
 \hat{\E} \left[ 
 {\bf 1}_{\{ \hat{P}_T (\mu) > \kappa \} } 
 \frac{\hat{P}_T({y}) }{\hat{P}_t(y) } 
 \Big| 
 {\cal F}_t \right] 
 \nu ( d {y} ) 
\\ 
 & = & 
 \hat{\E} \left[ 
 {\bf 1}_{\{ \hat{P}_T (\mu) > \kappa \} } 
 \frac{\hat{P}_T({y}) }{\hat{P}_t(y)  } 
 \Big| 
 {\cal F}_t \right] 
 ( 
 \mu ( d{y} ) 
 - 
 \kappa 
 \nu ( d {y} ) 
 ). 
\end{eqnarray*} 
\subsubsection*{Bond options} 
 In the case of a bond call option with $S=T$ and payoff 
 $(P_T(U) - \kappa )^+$ and $\mu = \delta_U$, $\nu = \delta_T$, 
 this yields 
\begin{equation} 
\label{cnc0} 
 \phi_t ( d{y} ) 
 = 
 \frac{{P}_t(T) }{{P}_t(U)} 
 \hat{\E} \left[ 
 {\bf 1}_{\{ \hat{P}_T ( U ) > \kappa \} } 
 \hat{P}_T ( U ) 
 \Big| 
 {\cal F}_t \right] 
 \delta_U ( d{y} ) 
 - \kappa 
 \hat{\E} \left[ 
 {\bf 1}_{\{ \hat{P}_T ( U ) > \kappa \} } 
 \Big| 
 {\cal F}_t \right] 
 \delta_T ( d {y} ) 
. 
\end{equation} 
 This particular setting of bond options can be 
 modeled using the diffusions of Section~\ref{mc} 
 since in that case 
 $\hat{P}_t (\mu) = P_t( U )/P_t( T )$ 
 is a geometric Brownian motion under $\hat{\P}$ 
 with volatility 
\begin{equation} 
\label{djlddaa} 
\hat{\sigma}(t)=\zeta _{t}( U )-\zeta _{t}( T )
\end{equation} 
 given by \eqref{asfvg} below, 
 in which case the above result coincides with the delta hedging 
 formula \eqref{cnc} below. 
\subsubsection*{Caplets on the LIBOR rate} 
 In the case of a caplet with payoff 
\begin{equation} 
\label{stl} 
 (S-T)(L(T,T,S)-\kappa )^{+}=(P_{T}(S)^{-1}-(1+\kappa (S-T)))^{+}, 
\end{equation} 
 on the LIBOR rate 
\begin{equation} 
\label{lts} 
 L(t,T,S) 
 = 
 \frac{P_t(T) - P_t(S)}{(S-T) P_t(S)} 
, 
 \qquad 
 0 \leq t \leq T < S, 
\end{equation} 
 and $\mu = \delta_T$, $\nu = \delta_S$, 
 Proposition~\ref{p52} yields 
\begin{eqnarray}
\label{dklddds.1} 
\phi _{t}(dy) &=&\frac{P_{t}(S)}{P_{t}(T)}\hat{\E} 
\left[ 
 \frac{1}{P_{T}(S)} 
 {\bf 1}_{\{P_{T}(S) < 1 / ( 1+\kappa (S-T) ) \}} 
\Big|\mathcal{F}_{t}\right] \delta _{T}(dy) 
\\
\nonumber 
&&-(1+\kappa (S-T))\hat{\E} 
\left[ \mathbf{1}_{\{P_{T}(S) < 1 / ( 1+\kappa (S-T) ) 
 \}}\Big| \mathcal{F}_{t}\right] \delta _S (d{y})
\end{eqnarray}
 In this case, $\hat{P}_t (\mu) = P_t(T)/P_t(S)$ 
 is modeled by a geometric 
 Brownian motion with volatility 
$ 
\hat{\sigma}(t)=\zeta _{t}( T )-\zeta _{t}( S )
$ 
 as in Section~\ref{mc} 
 and the above result coincides with the 
 formula \eqref{dklddds} below. 
\subsubsection*{Swaptions} 
 In this case the modeling of the swap rate differs 
 from the diffusion model of Section~\ref{mc}. 
 For a swaption with $S=T$ and payoff 
 $(P_T(T_i) - P_T(T_j) - \kappa P_T(\nu))^+$ 
 on the LIBOR, where 
$$ 
 \mu (d{y}) 
 = 
 \delta_{T_i} (d{y}) - \delta_{T_j} (d{y}) 
 \quad 
 \mbox{and} 
 \quad 
 \nu (d{y}) =  
 \sum_{k=i}^{j-1} 
 \tau_k 
 \delta_{T_{k+1}} (d{y}) 
, 
$$ 
 with $\tau_k = T_{k+1}-T_k$, $k=i,\ldots , j-1$, we obtain 
\begin{eqnarray} 
\nonumber 
 \phi_t ( d{y} ) 
 & = & 
 \hat{\E} \left[ 
 {\bf 1}_{\{ \hat{P}_T (\mu) > \kappa \} } 
 \frac{\hat{P}_{T_i} (T_i )}{ \hat{P}_t(T_i) } 
 \Big| 
 {\cal F}_t \right] 
 \delta_{T_i} ( d{y} ) 
 - 
 ( 1 + \kappa \tau_{j-1} ) 
 \hat{\E} \left[ 
 {\bf 1}_{\{ \hat{P}_T (\mu) > \kappa \} } 
 \frac{\hat{P}_{T_i}(T_j)}{\hat{P}_t (T_j ) } 
 \Big| 
 {\cal F}_t \right] 
 \delta_{T_j} ( d{y} ) 
\\ 
\label{dkldd111} 
 & & 
 - \kappa 
 \sum_{k=i+1}^{j-1} 
 \tau_{k-1} 
 \hat{\E} \left[ 
 {\bf 1}_{\{ \hat{P}_T (\mu) > \kappa \} } 
 \frac{\hat{P}_{T_i}(T_k) }{\hat{P}_t (T_k ) } 
 \Big| 
 {\cal F}_t \right] 
 \delta_{T_k} ( d {y} ) 
. 
\end{eqnarray} 
 The above consequence of Proposition~\ref{p52} below differs 
 from \eqref{sw} in Section~\ref{mc} because of different modeling 
 assumptions, as the deterministic volatility 
 \eqref{djlddaa} has no analog here, cf. \eqref{cs}, \eqref{wsht} 
 below. 
\begin{Proofy} \hskip-0.1cm {\em of Proposition~\ref{p52}}. 
 By Lemma~\ref{l2b} below the forward claim price 
$ 
 \hat{V}_t 
$ 
 has the predictable representation 
$$ 
 \hat{V}_t = 
 \hat{\E} [ \hat{\xi} ] 
 + 
 \int_0^t \langle {}\phi_s , d\hat{P}_s \rangle_{G^* \! ,G} 
, \qquad 
 0 \leq t \leq T. 
$$ 
 Hence by Lemma~\ref{p02} the portfolio priced as 
$$ 
 V_t 
 = 
 \langle 
 {}\phi_t 
 , 
 P_t 
 \rangle_{G^* \! ,G} 
, 
 \qquad 
 0 \leq t \leq T, 
$$ 
 is self-financing and it hedges the claim 
$ 
 \xi 
 = 
 P_S (\nu) \hat{g} 
 ( P_T(\mu) / P_T(\nu) )
$, 
 since $\eta_t = 0$ by \eqref{dw} and \eqref{bc}. 
\end{Proofy} 
\noindent 
 The next lemma, which will be used in the proof of Lemma~\ref{p1} 
 below, shows in particular that for fixed $U>0$, 
 $(\hat{P}_t(U))_{t\in \real_+}$ is usually not a geometric Brownian motion, 
 except in the case of bond options with 
 $\mu (d{y}) = \delta_U (d{y})$ and $\nu ( d{y} ) = \delta_T (d{y})$, 
 where we get 
$$ 
 d \frac{P_t(U)}{P_t(T)} 
 = 
 \frac{P_t(U)}{P_t(T)} 
 ( 
 \zeta_t(U) 
 - 
 \zeta_t(T) 
 ) 
 d\hat{W}_t 
, 
$$ 
 and 
\begin{equation} 
\label{asfvg} 
 \hat{\sigma} (t) 
 = 
 \zeta_t(U) 
 - 
 \zeta_t(T) 
, 
 \qquad 
 0 \leq t \leq T. 
\end{equation} 
\begin{lemma} 
\label{l} 
 For all ${y}\in \real_+$ we have  
$$ 
 d 
 \hat{P}_t ( y ) 
 = 
 \hat{\sigma}_t ( \hat{P}_t , y ) 
 d\hat{W}_t 
, 
 \qquad 
 t, y \in \real_+, 
$$ 
 where 
\begin{equation} 
\label{cs} 
 \hat{\sigma}_t ( \hat{P}_t , y ) 
 : = \hat{P}_t ( y ) 
 \int_T^\infty 
 \hat{P}_t ( z ) 
 ( 
 \zeta_t({y}) 
 - 
 \zeta_t({z}) 
 ) 
 \nu (d{z} ) 
, 
 \qquad 
 t, y \in \real_+. 
\end{equation} 
\end{lemma} 
\begin{Proof} 
 Defining the discounted bond price $\tilde{P}_t$ by 
\begin{equation} 
\label{tpt} 
 \tilde{P}_t = \exp \left( - \int_0^t r_s ds \right) P_t, 
 \qquad 
 t \in \real_+, 
\end{equation} 
 we have 
\begin{eqnarray*} 
 d \hat{P}_t(y) 
 & = & 
 d \left( \frac{\tilde{P}_t(y) }{\tilde{P}_t(\nu) } \right) 
\\ 
 & = & 
 \frac{d \tilde{P}_t(y)  }{\tilde{P}_t(\nu) } 
 + 
 \tilde{P}_t(y)  
 d \left( 
 \frac{1}{\tilde{P}_t(\nu) } 
 \right) 
 + 
 d \tilde{P}_t(y)  
 \cdot 
 d \left( 
 \frac{1}{\tilde{P}_t(\nu) } 
 \right) 
\\ 
 & = & 
 \frac{d \tilde{P}_t(y)  }{\tilde{P}_t(\nu) } 
 + 
 \frac{ \tilde{P}_t(y) }{\tilde{P}_t(\nu)} 
 \left( 
 - 
 \frac{d\tilde{P}_t(\nu) }{\tilde{P}_t(\nu)} 
 + 
 \left( 
 \frac{d\tilde{P}_t(\nu)}{\tilde{P}_t(\nu)} 
 \right)^2 
 \right) 
 - 
 \frac{ d \tilde{P}_t(y) }{\tilde{P}_t(\nu)} 
 \cdot 
 \frac{ d\tilde{P}_t(\nu) }{\tilde{P}_t(\nu)} 
\\ 
 & = & 
 \frac{ d \tilde{P}_t(y) }{\tilde{P}_t(\nu) } 
 - 
 \hat{P}_t(y) 
 \frac{d\tilde{P}_t(\nu)  }{\tilde{P}_t(\nu)} 
\\ 
 & & 
 + 
 \hat{P}_t(y) 
 \int_T^\infty 
 \hat{P}_t(s) 
 \int_T^\infty 
 \hat{P}_t(z) 
 \zeta_t(z) \zeta_t({s}) \nu (dz ) \nu (d{s} ) 
 dt 
\\ 
 & & 
 - 
 \zeta_t(y) 
 \hat{P}_t(y) 
 \int_T^\infty 
 \hat{P}_t( z ) 
 \zeta_t( z ) \nu (dz) dt 
\\ 
 & = & 
 \hat{P}_t(y) 
 \zeta_t(y) dW_t 
 - 
 \hat{P}_t(y) 
 \int_T^\infty 
 \hat{P}_t(z) 
 \zeta_t(z) 
 \nu (dz ) 
 dW_t 
\\ 
 & & 
 - 
 \hat{P}_t(y) 
 \int_T^\infty 
 \hat{P}_t(s) 
 \int_T^\infty 
 \hat{P}_t(z) 
 ( \zeta_t( y ) - \zeta_t(z) ) 
 \zeta_t({s}) \nu (dz ) \nu (d{s} ) 
 dt 
\\ 
 & = & 
 \hat{P}_t(y) 
 \int_T^\infty 
 \hat{P}_t(z) 
 ( 
 \zeta_t(y) 
 - 
 \zeta_t({z}) 
 ) 
 \nu (d{z} ) 
 dW_t 
\\ 
 & & 
 - 
 \hat{P}_t(y) 
 \int_T^\infty 
 \hat{P}_t(z) 
 ( \zeta_t( y ) - \zeta_t(z) ) 
 \int_T^\infty 
 \hat{P}_t(s) 
 \zeta_t({s}) \nu (d{s} ) \nu (dz ) 
 dt 
\\ 
 & = & 
 \hat{P}_t(y) 
 \int_T^\infty 
 \hat{P}_t(z) 
 ( 
 \zeta_t(y) 
 - 
 \zeta_t({z}) 
 ) 
 \nu (d{z} ) 
 d\hat{W}_t 
, 
\end{eqnarray*} 
 by the relation 
\begin{equation} 
\label{**3} 
 d \hat{W}_t 
 = 
 d W_t 
 - 
 \int_T^\infty 
 \hat{P}_t (s)
 \zeta_t ( {s} ) 
 \nu ( d{s} ) 
 dt 
, 
 \qquad 
 t \in \real_+ 
, 
\end{equation} 
 which follows from \eqref{*2}. 
\end{Proof} 
 In the case of a swaption with 
 $\mu (d{y}) 
 = 
 \delta_{T_i} (d{y}) - \delta_{T_j} (d{y})$ 
 and 
 $\nu (d{y}) =  
 \sum_{k=i}^{j-1} 
 \tau_k 
 \delta_{T_{k+1}} (d{y})$, 
 $\hat{P}_t (\mu )$ becomes the corresponding 
 swap rate and Lemma~\ref{l} yields 
$$ 
 d \frac{P_t(\mu)}{P_t(\nu)} 
 = 
 \frac{P_t(\mu)}{P_t(\nu)} 
 \left( 
 \frac{P_t(T_j)}{P_t(\mu)} 
 ( 
 \zeta_t({T_i}) 
 - 
 \zeta_t({T_j}) 
 ) 
 + 
 \sum_{k=i}^{j-1} 
 \tau_k 
 \frac{P_t(T_{k+1})}{P_t(\nu)} 
 ( 
 \zeta_t({T_i}) 
 - 
 \zeta_t({T_{k+1}}) 
 ) 
 \right) 
 d\hat{W}_t 
, 
$$ 
 which shows that 
\begin{equation} 
\label{wsht} 
 \hat{\sigma} (t) 
 = 
 \frac{P_t(T_j)}{P_t(\mu)} 
 ( 
 \zeta_t({T_i}) 
 - 
 \zeta_t({T_j}) 
 ) 
 + 
 \sum_{k=i}^{j-1} 
 \tau_k 
 \frac{P_t(T_{k+1})}{P_t(\nu)} 
 ( 
 \zeta_t({T_i}) 
 - 
 \zeta_t({T_{k+1}}) 
 ) 
, 
\end{equation} 
 $0 \leq t \leq T$, 
 and coincides with the dynamics of the LIBOR swap rate 
 in Relation~(1.28), page 17 of \cite{schoenmakersbk}. 
\\ 
 
 Lemma~\ref{nl} has been used in the proof of Proposition~\ref{p52}. 
\begin{lemma} 
\label{nl} 
 We have 
\begin{equation} 
\label{ds} 
 D_t \hat{P}_u (y) 
 = 
 \hat{\sigma}_t ( \hat{P}_u , y ) 
, 
 \qquad 
 0 \leq t \leq u, 
 \quad 
 y \in \real_+, 
\end{equation} 
 where 
\begin{equation} 
\label{jklff} 
 \hat{\sigma}_t ( \hat{P}_u , y ) 
 = 
 \hat{P}_u(y) 
 \int_T^\infty 
 \hat{P}_u(z) 
 ( 
 \zeta_t ( {y} ) 
 - 
 \zeta_t ( {z} ) 
 ) 
 \nu (d{z}) 
, 
\end{equation} 
 $0 \leq t \leq u$, $y \in \real_+$. 
\end{lemma} 
\begin{Proof} 
 The discounted bond price $\tilde{P}_t$ defined in 
 \eqref{tpt} satisfies the relation 
$$ 
 \tilde{P}_u(y) = 
 \tilde{P}_0(y) 
 \exp \left( \int_0^u \zeta_t ({y}) dW_t 
 - 
 \frac{1}{2} 
 \int_0^u 
 \left\vert 
 \zeta_t({y}) 
 \right\vert^2 
 dt\right), 
 \qquad 
 {y} \in \real_+, 
$$ 
 with 
$$ 
 D_u 
 \tilde{P}_T (y) 
 = 
 \tilde{P}_T (y) 
 \zeta_u({y}), 
 \qquad 
 0 \leq u \leq T, 
 \quad {y} \in \real_+ 
. 
$$ 
 Hence from the relation 
$$ 
 d\tilde{P}_u(y) = \zeta_u ({y} ) \tilde{P}_u(y) dW_t 
, 
 \qquad 
 {y} \in \real_+, 
$$ 
 we get 
\begin{eqnarray*}
 D_t \hat{P}_u(y) 
 & = & 
 D_t \frac{ \tilde{P}_u(y)}{\tilde{P}_u(\nu) } 
\\ 
 & = & 
 \frac{ D_t \tilde{P}_u(y) }{\tilde{P}_u(\nu) } 
 - 
 \frac{\tilde{P}_u(y)}{\tilde{P}_u(\nu) } 
 \frac{ D_t \tilde{P}_u(\nu) }{\tilde{P}_u(\nu) } 
\\ 
 & = & 
 \frac{\tilde{P}_u(y)}{\tilde{P}_u(\nu) } 
 \left( 
 \zeta_t( {y} ) 
 - 
 \int_T^\infty 
 \zeta_t ( {z} ) 
 \frac{ \tilde{P}_u(z) }{\tilde{P}_u(\nu) } 
 \nu (d{z}) 
 \right) 
\\ 
 & = & 
 \hat{P}_u(y) 
 \int_T^\infty 
 \hat{P}_u(z) 
 ( 
 \zeta_t ( {y} ) 
 - 
 \zeta_t ( {z} ) 
 ) 
 \nu (d{z}) 
\\ 
 & = & 
 \hat{\sigma}_t ( \hat{P}_u , y ) 
, 
\end{eqnarray*} 
 $0 \leq t \leq u$, $y \in \real_+$. 
\end{Proof} 
 The following lemma has been used in the proof of Lemma~\ref{l2b}. 
\begin{lemma} 
\label{p1}
 Taking $\hat{\xi} = \hat{g} ( \hat{P}_T(\mu ) )$, 
 the process in Lemma~\ref{cf} 
 is given by 
\begin{eqnarray*} 
\nonumber 
 \hat{\alpha}_t 
 & = & 
 \int_T^\infty 
 \hat{\E} \left[ 
 \hat{g}' ( \hat{P}_T(\mu ) ) 
 \hat{P}_T (y) 
 \Big| 
 {\cal F}_t \right] 
 \zeta_t ( {y} ) 
 \mu ( d{y} ) 
\\ 
 & & 
 - 
 \int_T^\infty 
 \hat{\E} \left[ 
 \hat{P}_T(\mu ) 
 \hat{g}' ( \hat{P}_T(\mu ) ) 
 \hat{P}_T(y) 
 \Big| 
 {\cal F}_t \right] 
 \zeta_t ( {y} ) 
 \nu (d{y}) 
\\
\nonumber 
 & & 
 + 
 \int_T^\infty 
 \hat{\E} \left[ 
 \hat{g} ( \hat{P}_T(\mu ) ) 
 ( 
 \hat{P}_T (y) 
 - 
 \hat{P}_t(y) 
 ) 
 \Big| 
 {\cal F}_t \right] 
 \zeta_t ( {y} ) 
 \nu ( dy ) 
\end{eqnarray*} 
\end{lemma} 
\begin{Proof} 
 By \eqref{**3}, 
 the process $(\gamma_t)_{t\in \real_+}$ in 
 \eqref{adfdsg} is given by 
$$ 
 \gamma_t = 
 \int_T^\infty 
 \hat{P}_t (s)
 \zeta_t ( {s} ) 
 \nu ( d{s} ) \in H
, 
 \qquad 
 t \in \real_+. 
$$ 
 Taking $\hat{\xi} = \hat{g} ( \hat{P}_T(\mu ) )$, 
 Lemma~\ref{cf} yields 
$$ 
 \hat{V}_t = \hat{\E} [ \hat{g} ( \hat{P}_T(\mu ) ) ] 
 + 
 \int_0^t \langle \hat{\alpha}_s , d\hat{W}_s \rangle_H 
, \qquad 
 0 \leq t \leq T, 
$$ 
 where 
\begin{equation} 
\label{fda} 
 \hat{\alpha}_s = 
 \hat{\E} \left[ 
 D_s \hat{g} ( \hat{P}_T(\mu ) ) 
 + 
 \hat{g} ( \hat{P}_T(\mu ) ) 
 \int_s^T D_s 
 \int_T^\infty 
 \hat{P}_u (y)
 \zeta_u ( {y} ) 
 \nu ( d{y} ) 
 d\hat{W}_u 
 \Big| 
 {\cal F}_s 
 \right] 
, 
\end{equation}  
 $0 \leq s \leq T$. 
 By integration with respect to $\mu ( dy )$ in \eqref{ds} we get 
$$ 
 D_t \hat{P}_T(\mu ) 
 = 
 \int_T^\infty 
 \zeta_t ( {y} ) 
 \hat{P}_T({y}) 
 \mu ( d{y} ) 
 - 
 \hat{P}_T(\mu ) 
 \int_T^\infty 
 \zeta_t ( {y} ) 
 \hat{P}_T({y}) 
 \nu (d{y}) 
, 
$$ 
 which allows us to compute 
 $D_t \hat{g} ( \hat{P}_T(\mu ) ) 
 = 
 \hat{g}' 
 ( \hat{P}_T(\mu ) ) 
 D_t \hat{P}_T(\mu )$ in \eqref{fda}, 
 $0 \leq t \leq T$. 
 On the other hand, by Lemmas~\ref{l} and \ref{nl} 
 the second term in \eqref{fda} can be computed as 
\begin{eqnarray*} 
\lefteqn{ 
\! \! \! \! \! \! \! \! \! \! \! \! \! \! \! \! \! \! 
\! \! \! \! \! \! \! \! \! \! \! \! \! \! \! \! \! \! 
\! \! \! \! \! \! 
 \int_t^T 
 D_t 
 \int_T^\infty 
 \hat{P}_u (y) 
 \zeta_u ( {y} ) 
 \nu ( d{y} ) 
 d\hat{W}_u 
 = 
 \int_t^T 
 \int_T^\infty 
 \hat{\sigma}_t ( \hat{P}_u , y ) 
 \zeta_u ( {y} ) 
 \nu ( d{y} ) 
 d\hat{W}_u 
} 
\\ 
 & = & 
 \int_t^T 
 \int_T^\infty 
 \hat{\sigma}_u ( \hat{P}_u , y ) 
 \zeta_t ( {y} ) 
 \nu ( d{y} ) 
 d\hat{W}_u 
\\ 
 & = & 
 \int_T^\infty 
 \int_t^T 
 \hat{\sigma}_u ( \hat{P}_u , y ) 
 d\hat{W}_u 
 \zeta_t ( {y} ) 
 \nu ( d{y} ) 
\\ 
 & = & 
 \int_T^\infty 
 \int_t^T 
 d 
 \hat{P}_u (y) 
 \zeta_t ( {y} ) 
 \nu ( d{y} ) 
\\ 
 & = & 
 \int_T^\infty 
 ( 
 \hat{P}_T (y) 
 - 
 \hat{P}_t(y) 
 ) 
 \zeta_t ( {y} ) 
 \nu ( d{y} ) 
, 
\end{eqnarray*} 
 where $\hat{\sigma}_t ( \hat{P}_u , y )$ 
 is given by \eqref{jklff} above, hence 
\begin{eqnarray*} 
\lefteqn{ 
 D_t 
 \hat{g} ( \hat{P}_T(\mu ) ) 
 + 
 \hat{g} ( \hat{P}_T(\mu ) ) 
 \int_t^T 
 D_t 
 \int_T^\infty 
 \hat{P}_u(y) 
 \zeta_t ( {y} ) 
 \nu ( d{y} ) 
 d\hat{W}_u 
} 
\\
&=& 
 \hat{g}' ( \hat{P}_T(\mu ) ) 
 D_t \hat{P}_T(\mu ) 
 + 
 \hat{g} ( \hat{P}_T(\mu ) ) 
 \int_t^T 
 D_t 
 \int_T^\infty 
 \hat{P}_u(y) 
 \zeta_t ( {y} ) 
 \nu ( d{y} ) 
 d\hat{W}_u 
\\ 
 &=& 
 \hat{g}' ( \hat{P}_T(\mu ) ) 
 \int_T^\infty 
 \zeta_t ( {y} ) 
 \hat{P}_T (y) 
 \mu ( d{y} ) 
 - 
 \hat{P}_T(\mu ) 
 \hat{g}' ( \hat{P}_T(\mu ) ) 
 \int_T^\infty 
 \zeta_t ( {y} ) 
 \hat{P}_T(y) 
 \nu (d{y}) 
\\
 & & 
 + 
 \int_T^\infty 
 \hat{g} ( \hat{P}_T(\mu ) ) 
 ( 
 \hat{P}_T (y) 
 - 
 \hat{P}_t(y) 
 ) 
 \zeta_t ( {y} ) 
 \nu ( dy ) 
, 
\end{eqnarray*} 
 which is square-integrable by Conditions~\eqref{1.k} and \eqref{2.k}. 
\\ 
 
 By \eqref{fda}, this yields 
\begin{eqnarray*} 
\nonumber 
 \hat{\alpha}_t 
 & = & 
 \int_T^\infty 
 \hat{\E} \left[ 
 \hat{g}' ( \hat{P}_T(\mu ) ) 
 \hat{P}_T (y) 
 \Big| 
 {\cal F}_t \right] 
 \zeta_t ( {y} ) 
 \mu ( d{y} ) 
\\ 
 & & 
 - 
 \int_T^\infty 
 \hat{\E} \left[ 
 \hat{P}_T(\mu ) 
 \hat{g}' ( \hat{P}_T(\mu ) ) 
 \hat{P}_T(y) 
 \Big| 
 {\cal F}_t \right] 
 \zeta_t ( {y} ) 
 \nu (d{y}) 
\\
\nonumber 
 & & 
 + 
 \int_T^\infty 
 \hat{\E} \left[ 
 \hat{g} ( \hat{P}_T(\mu ) ) 
 ( 
 \hat{P}_T (y) 
 - 
 \hat{P}_t(y) 
 ) 
 \Big| 
 {\cal F}_t \right] 
 \zeta_t ( {y} ) 
 \nu ( dy ) 
\end{eqnarray*} 
\end{Proof} 
 The next lemma has been used in the proof of Proposition~\ref{p52}. 
\begin{lemma} 
\label{l2b} 
 The process ${}\phi_t$ in 
 the predictable representation 
$$ 
 \hat{V}_t = 
 \hat{\E} [ \hat{\xi} ] 
 + 
 \int_0^t \langle {}\phi_s , d\hat{P}_s \rangle_{G^* \! ,G} 
, \qquad 
 0 \leq t \leq T, 
$$ 
 of the forward claim price 
$ 
 \hat{V}_t 
 : =  \hat{\E} 
 [ 
 \hat{\xi} 
 | 
 {\cal F}_t 
 ] 
$, 
 cf. \eqref{prd}, is given by 
$$ 
 \phi_t ( d{y} ) 
 = 
 \hat{\E} \left[ 
 \frac{\hat{P}_T({y}) }{\hat{P}_t( y )  } 
 \hat{g}' ( \hat{P}_T(\mu ) )  
 \Big| 
 {\cal F}_t \right] 
 \mu ( d{y} ) 
 + 
 \hat{\E} \left[ 
 ( 
 \hat{g} ( \hat{P}_T(\mu ) )  
 - 
 \hat{P}_T(\mu ) 
 \hat{g}' ( \hat{P}_T(\mu ) )  
 ) 
 \frac{\hat{P}_T({y}) }{\hat{P}_t( y ) } 
 \Big| 
 {\cal F}_t \right] 
 \nu ( d {y} ) 
, 
$$ 
 $0\leq t \leq T$, 
\end{lemma} 
\begin{Proof} 
 By Lemma~\ref{p1} above we have, since 
 $\displaystyle 
 \hat{P}_t(\nu) 
 = 
 \int_T^\infty \frac{P_t(y)}{P_t(\nu)} \nu (dy ) = 1$,  
\begin{eqnarray} 
\nonumber 
 \langle \hat{\alpha}_t , dW_t \rangle_H 
 & = & 
 \int_T^\infty 
 \hat{\E} \left[ 
 \hat{g}' ( \hat{P}_T(\mu ) ) 
 \hat{P}_T (y) 
 \Big| 
 {\cal F}_t \right] 
 \zeta_t ( {y} ) 
 \mu ( d{y} ) 
 dW_t 
\\ 
\nonumber 
 & & 
 - 
 \int_T^\infty 
 \hat{\E} \left[ 
 \hat{P}_T(\mu ) 
 \hat{g}' ( \hat{P}_T(\mu ) ) 
 \hat{P}_T(y) 
 \Big| 
 {\cal F}_t \right] 
 \zeta_t ( {y} ) 
 \nu (d{y}) 
 dW_t 
\\ 
\nonumber 
 & & 
 + 
 \int_T^\infty 
 \hat{\E} \left[ 
 \hat{g} ( \hat{P}_T(\mu ) ) 
 ( 
 \hat{P}_T (y) 
 - 
 \hat{P}_t(y) 
 ) 
 \Big| 
 {\cal F}_t \right] 
 \zeta_t ( {y} ) 
 \nu ( dy ) 
 dW_t 
\\ 
\nonumber 
 & = & 
 \int_T^\infty 
 \hat{\E} \left[ 
 \hat{g}' ( \hat{P}_T(\mu ) ) 
 \hat{P}_T (y) 
 \Big| 
 {\cal F}_t \right] 
 \mu ( d{y} ) 
 \left( \frac{d{P}_t(y)}{{P}_t(y)} - r_t dt \right) 
\\ 
\nonumber 
 & & 
 - 
 \int_T^\infty 
 \hat{\E} \left[ 
 \hat{P}_T(\mu ) 
 \hat{g}' ( \hat{P}_T(\mu ) ) 
 \hat{P}_T(y) 
 \Big| 
 {\cal F}_t \right] 
 \nu (d{y}) 
 \left( \frac{d{P}_t(y)}{{P}_t(y)} - r_t dt \right) 
\\ 
\nonumber 
 & & 
 + 
 \int_T^\infty 
 \hat{\E} \left[ 
 \hat{g} ( \hat{P}_T(\mu ) ) 
 ( 
 \hat{P}_T (y) 
 - 
 \hat{P}_t(y) 
 ) 
 \Big| 
 {\cal F}_t \right] 
 \nu ( dy ) 
 \left( \frac{d{P}_t(y)}{{P}_t(y)} - r_t dt \right) 
\\ 
\nonumber 
 & = & 
 \int_T^\infty 
 \hat{\E} \left[ 
 \hat{g}' ( \hat{P}_T(\mu ) ) 
 \hat{P}_T (y) 
 \Big| 
 {\cal F}_t \right] 
 \mu ( d{y} ) 
 \frac{d{P}_t(y)}{{P}_t(y)} 
\\ 
\nonumber 
 & & 
 - 
 \int_T^\infty 
 \hat{\E} \left[ 
 \hat{P}_T(\mu ) 
 \hat{g}' ( \hat{P}_T(\mu ) ) 
 \hat{P}_T(y) 
 \Big| 
 {\cal F}_t \right] 
 \nu (d{y}) 
 \frac{d{P}_t(y)}{{P}_t(y)} 
\\ 
\nonumber 
 & & 
 + 
 \int_T^\infty 
 \hat{\E} \left[ 
 \hat{g} ( \hat{P}_T(\mu ) ) 
 ( 
 \hat{P}_T (y) 
 - 
 \hat{P}_t(y) 
 ) 
 \Big| 
 {\cal F}_t \right] 
 \nu ( dy ) 
 \frac{d{P}_t(y)}{{P}_t(y)} 
\\ 
\nonumber 
 & = & 
 \int_T^\infty 
 \hat{\E} \left[ 
 \hat{P}_T (y) 
 \hat{g}' ( \hat{P}_T(\mu ) )  
 \Big| 
 {\cal F}_t \right] 
 \mu ( d{y} ) 
 \frac{d{P}_t(y)}{{P}_t(y)} 
\\ 
\nonumber 
 & & 
 + 
 \int_T^\infty 
 \hat{\E} \left[ 
 \hat{P}_T (y) 
 ( 
 \hat{g} ( \hat{P}_T(\mu ) )  
 - 
 \hat{P}_T(\mu ) 
 \hat{g}' ( \hat{P}_T(\mu ) )  
 ) 
 \Big| 
 {\cal F}_t \right] 
 \nu (d{y}) 
 \frac{d{P}_t(y)}{{P}_t(y)} 
\\ 
\nonumber 
 & & 
 - 
 \hat{\E} \left[ 
 \hat{g} ( \hat{P}_T(\mu ) )  
 \Big| 
 {\cal F}_t \right] 
 \frac{ d{P}_t(\nu) }{{P}_t(\nu) } 
\\ 
\nonumber 
 & = & 
 \frac{1}{M_t} 
 \langle \phi_t , d{P}_t(y) \rangle_{G^* \! ,G} 
 - 
 \hat{V}_t 
 \frac{ d{P}_t(\nu) }{{P}_t(\nu) } 
, 
\end{eqnarray} 
 and by \eqref{*2} and \eqref{bc} we have 
\begin{eqnarray} 
\nonumber 
 \langle \hat{\alpha}_t , d\hat{W}_t \rangle_H 
 & = & 
 \langle \hat{\alpha}_t , d{W}_t \rangle_H 
 - 
 \frac{1}{M_t} 
 dM_t \cdot \langle \hat{\alpha}_t , d {W}_t \rangle_H 
\\ 
\nonumber 
 & = & 
 \langle \hat{\alpha}_t , d{W}_t \rangle_H 
 - 
 \frac{1}{M_t} 
 dM_t 
 \cdot 
 \left( 
 \frac{1}{M_t} \langle \phi_t , dP_t \rangle_{G^* \! ,G} 
 - 
 \frac{1}{M_t} \hat{V}_t dM_t 
 \right) 
\\ 
\nonumber 
 & = & 
 \langle \hat{\alpha}_t , d{W}_t \rangle_H 
 - 
 \frac{1}{M_t} 
 dM_t 
 \cdot 
 \left( 
 \langle \phi_t , d\hat{P}_t \rangle_{G^* \! ,G} 
 + 
 \frac{1}{M_t} \langle \phi_t , \hat{P}_t \rangle_{G^* \! ,G} dM_t 
\right. 
\\ 
\nonumber 
 & & 
\left. 
 + 
 \frac{1}{M_t} dM_t \cdot \langle \phi_t , d\hat{P}_t \rangle_{G^* \! ,G} 
 - 
 \frac{1}{M_t} \hat{V}_t dM_t 
 \right) 
\\ 
\nonumber 
 & = & 
 \langle \hat{\alpha}_t , d{W}_t \rangle_H 
 - 
 \frac{1}{M_t} 
 dM_t 
 \cdot 
 \left( 
 \langle \phi_t , d\hat{P}_t \rangle_{G^* \! ,G} 
 + 
 \frac{1}{M_t} dM_t \cdot \langle \phi_t , d\hat{P}_t \rangle_{G^* \! ,G} 
 \right) 
\\ 
\nonumber 
 & = & 
 \frac{1}{M_t} \langle \phi_t , dP_t \rangle_{G^* \! ,G} 
 - 
 \frac{1}{M_t} \hat{V}_t dM_t 
 - 
 \frac{1}{M_t} 
 dM_t 
 \cdot 
 \langle \phi_t , d\hat{P}_t \rangle_{G^* \! ,G} 
\\ 
\label{from} 
 & = & 
 \langle \phi_t , d\hat{P}_t \rangle_{G^* \! ,G} 
, 
\end{eqnarray} 
 since 
$$
 dP_t 
 = 
 M_t d\hat{P}_t 
 + 
 \hat{V}_t dM_t 
 + 
 dM_t 
 \cdot 
 d\hat{P}_t 
. 
$$ 
\end{Proof} 
 When the forward price process $(\hat{P}_t)_{t\in \real_+}$ 
 follows the dynamics \eqref{isc}, 
 Relation~\eqref{from} above shows that we have the relation 
$$ 
 \langle \hat{\alpha}_t , d\hat{W}_t \rangle_H 
 = 
 \langle \phi_t , d\hat{P}_t \rangle_{G^* \! ,G} 
 = 
 \langle \phi_t , \hat{\sigma}_t d\hat{W}_t \rangle_{G^* \! ,G} 
, 
$$ 
 which shows that 
$$ 
 \hat{\alpha}_t 
 = 
 \hat{\sigma}^*_t \phi_t 
, 
$$ 
 and recovers \eqref{asfghfsd}. 
\section{Delta hedging} 
\label{mc} 
 In this section we consider a $G$-valued 
 asset price process $(X_t)_{t\in \real_+}$ 
 and a numeraire $(M_t)_{t\in \real_+}$, 
 and we assume that the forward asset price 
 $\hat{X}_t : = \hat{X}_t /M_t$, $t\in \real_+$, 
 is modeled by the diffusion equation 
\begin{equation} 
\label{mp} 
 d \hat{X}_t = \hat{\sigma}_t ( \hat{X}_t ) d\hat{W}_t, 
\end{equation} 
 under the forward measure $\hat{\P}$ defined by 
 \eqref{fwd}, 
 where $x \mapsto \hat{\sigma}_t ( x ) \in {\cal L}_{HS} ( H , G )$ 
 is a Lipschitz function from $G$ 
 into the space of Hilbert-Schmidt operators from $H$ to $G$, 
 uniformly in $t\in \real_+$, 
\subsubsection*{Vanilla options} 
 In this Markovian setting a Vanilla option 
 with payoff $ 
 \xi = 
 M_S 
 \hat{g} 
 ( 
 \hat{X}_T 
 ) 
$ 
 is priced at time $t$ as 
\begin{equation} 
\label{eq01} 
 \E\left[ 
 e^{-\int_t^S r_s ds} 
 M_S 
 \hat{g} 
 ( 
 \hat{X}_T 
 ) 
 \Big| 
 {\cal F}_t 
 \right] 
 = 
 M_t 
 \hat{\E} 
 \left[ 
 \hat{g} 
 ( 
 \hat{X}_T 
 ) 
 \Big| 
 {\cal F}_t 
 \right] 
 = 
 M_t 
 \hat{C} ( t , \hat{X}_t ) 
, 
\end{equation} 
 for some measurable function $\hat{C} (t,x)$ on $\real_+ \times G$, 
 and Lemma~\ref{p02} has the following corollary. 
\begin{corollary} 
\label{p1.1} 
 Assume that the function $\hat{C} (t,x)$ 
 is ${\cal C}^2$ on $\real_+ \times G$, and let 
$${}\eta_t = 
 \hat{C} ( t , \hat{X}_t ) 
 - \langle  \nabla \hat{C} ( t , \hat{X}_t ) , \hat{X}_t \rangle_{G^* \! ,G}, 
 \qquad 
 0 \leq t \leq T. 
$$ 
 Then 
 the portfolio $( \nabla \hat{C} ( t , \hat{X}_t ) ,{}\eta_t)_{t\in [0,T]}$ 
 with value 
$$ 
 V_t 
 = 
 {}\eta_t  
 M_t 
 + 
 \langle 
 \nabla \hat{C} ( t , \hat{X}_t ) 
 , 
 X_t 
 \rangle_{G^* \! ,G} 
, 
 \qquad 
 0 \leq t \leq T, 
$$ 
 is self-financing and hedges the claim $\xi = M_S \hat{g} (\hat{X}_T)$. 
\end{corollary} 
\begin{Proof} 
 By It\^o's formula, cf. Theorem~4.17 of \cite{daprato}, 
 and the martingale property of $\hat{V}_t$ under $\hat{\P}$, 
 the predictable representation \eqref{prd} is given by 
$$ 
 {}\phi_t = 
 \nabla \hat{C} ( t , \hat{X}_t )
, 
 \qquad 
 0 \leq t \leq T. 
$$ 
\end{Proof} 
 When 
$$ 
 X_t = P_t(\mu) 
 : = 
 \langle \mu , P_t \rangle_{G^* \! ,G} 
 = 
 \int_T^\infty 
 P_t(y) \mu (d{y}) 
, 
$$ 
 and 
$$ 
 M_t = 
 P_t ( \nu ) = 
 \langle \nu , P_t \rangle_{G^* \! ,G} 
 = 
 \int_T^\infty 
 P_t(y) \nu (d{y}) 
, 
$$ 
 Corollary~\ref{p1.1} 
 shows that the portfolio 
\begin{equation} 
\label{cf1} 
 \phi_t ( d{y} ) 
 = 
 \frac{\partial \hat{C}}{\partial x} ( t , \hat{X}_t ) 
 \mu ( d{y} ) 
 + 
 \left( 
 \hat{C} ( t , \hat{X}_t ) 
 - 
 \hat{X}_t \frac{\partial \hat{C}}{\partial x} ( t , \hat{X}_t ) 
 \right) 
 \nu ( d{y} ) 
, 
\end{equation} 
 $0 \leq t \leq T$, 
 where $\hat{C} ( t , x )$ is defined in \eqref{eq01}, 
 is a self-financing hedging strategy for the claim 
$$ 
 \xi 
 = 
 P_S (\nu)  
 \hat{g} 
 \left(
 \frac{ 
 P_T(\mu) }{ 
 P_T(\nu)  } 
 \right) 
, 
$$ 
 with $M_t = P_t (\nu)$, $t \in \real_+$. 
\\ 
 
 When $G=\real$ and 
 $(\hat{X}_t)_{t\in \real_+}$ is a geometric Brownian motion 
 with deterministic volatility $H$-valued function 
 $( \hat{\sigma} (t) )_{t\in \real_+}$ 
 under the forward measure $\hat{\P}$, i.e. 
\begin{equation} 
\label{utfm} 
 d \hat{X}_t = \hat{X}_t \hat{\sigma}_t ( t) d\hat{W}_t, 
\end{equation} 
 the exchange call option with payoff 
$$ 
 M_S 
 ( 
 \hat{X}_T 
 - 
 \kappa 
 )^+ 
, 
$$ 
 is priced by 
 the Black-Scholes-Margrabe formula 
\begin{equation} 
\label{mg} 
 \E\left[ 
 e^{-\int_t^S r_s ds} 
 \left( 
 X_T 
 - 
 \kappa 
 M_T 
 \right)^+ 
 \Big| 
 {\cal F}_t 
 \right] 
 = 
 X_t 
 \Phi^0_+ ( t , \kappa , \hat{X}_t ) 
 - 
 \kappa 
 M_t 
 \Phi^0_- 
 ( t , \kappa , \hat{X}_t ) 
, 
 \qquad 
 t\in \real_+, 
\end{equation}  
 where 
\begin{equation} 
\label{phi} 
 \Phi^0_+ ( t , \kappa , x ) 
 = 
 \Phi 
 \left( 
 \frac{\log ( x / \kappa )}{ v (t,T) } 
 + 
 \frac{ v(t,T)}{2} 
 \right) 
 \quad 
 \mbox{ and } 
 \quad 
 \Phi^0_- (t , \kappa , x ) 
 = 
 \Phi 
 \left( 
 \frac{\log ( x / \kappa )}{ v (t,T) } 
 - 
 \frac{ v(t,T)}{2} 
 \right) 
, 
\end{equation} 
 and 
$$ 
 v^2(t,T) = 
 \int_{t}^T \hat{\sigma}^2 (s) ds 
. 
$$ 
 By Corollary~\ref{p1.1} and the relation 
$$ 
 \frac{\partial \hat{C}}{\partial x} ( t , x ) 
 = 
 \Phi \left( 
 \frac{ 
 \log
 ( x / \kappa  ) 
 }{ v (t,T) } 
 + \frac{v(t,T)}{2} 
 \right) 
 = 
 \Phi^0_+ ( t , \kappa , x ) 
, 
$$ 
 this yields a self-financing portfolio 
$$ 
 ( 
 \Phi^0_+ ( t , \kappa , \hat{X}_t ) , 
 - 
 \kappa 
 \Phi^0_- 
 ( t , \kappa , \hat{X}_t ) )_{t\in [0,T]} 
$$ 
 in $(X_t,M_t)$ that hedges the claim 
 $
 \xi = 
 \left( 
 X_T 
 - 
 \kappa 
 M_T 
 \right)^+$. 
 In particular, 
 when the short rate process $(r_t)_{t\in \real_+}$ is a deterministic 
 function and 
$ 
\displaystyle 
 M_t = e^{-\int_t^T r_s ds}
$, 
 $0 \leq t \leq T$, \eqref{mg} is Merton's 
 ``zero interest rate'' version 
 of the Black-Scholes formula, 
 a property which has been used in \cite{jamshidian2} 
 for the hedging of swaptions. 
\\ 
 
 In particular, from \eqref{mg} we have 
\begin{eqnarray} 
\label{bsf} 
 \E\left[ 
 e^{-\int_t^S r_s ds} 
 P_S(\nu ) ( \hat{X}_T - \kappa )^+ 
 \Big| 
 {\cal F}_t 
 \right] 
 & = & 
 P_t ( \nu ) 
 \hat{C} ( t , \hat{X}_t ) 
\\ 
\nonumber 
 & = & 
 P_t(\mu)  
 \Phi^0_+ ( t , \kappa , \hat{X}_t ) 
 - 
 \kappa 
 P_t(\nu) 
 \Phi^0_- ( t , \kappa , \hat{X}_t ) 
, 
\end{eqnarray} 
 and the portfolio 
\begin{equation} 
\label{c0a} 
 \phi_t ( d{y} ) 
 = 
 \Phi^0_+ ( t , \kappa , \hat{X}_t ) 
 \mu ( d{y} ) 
 - 
 \kappa 
 \Phi^0_- ( t , \kappa , \hat{X}_t ) 
 \nu ( d{y} ) 
, 
 \qquad 
 0 \leq t \leq T, 
\end{equation} 
 is self-financing, hedges the claim 
 $(P_T(\mu) - \kappa P_T(\nu))^+$, and is evenly distributed 
 with respect to $\mu (dy)$ and to $\nu (dy)$. 
\\ 
 
 As applications of \eqref{cf1} and \eqref{bsf}, 
 we consider some examples of delta hedging, in which the 
 asset allocation is uniform on $\mu (dy)$ and $\nu (dy)$ 
 with respect to the bond maturities $y\in [T,\infty )$. 
\subsubsection*{Bond options} 
 Taking $S=T$, the bond option with payoff 
$$ 
 \xi 
 = 
 M_T 
 \hat{g} 
 ( 
 P_T( U ) 
 ) 
, 
 \qquad 
 0 \leq T \leq  U, 
$$ 
 belongs to the above framework with 
$$ 
 \mu (d{y}) = \delta_U (d{y}) 
 \quad 
 \mbox{ and } 
 \quad 
 \nu ( d{y} ) = \delta_T (d{y}) 
, 
$$ 
 hence $M_t = P_t(\nu)  = P_t(T)$ 
 and when $\hat{X}_t = P_t(U)/P_t(T)$ is Markov as in 
 \eqref{mp}, 
 the self-financing hedging strategy is given from \eqref{cf1} by 
\begin{equation} 
\label{sfhs} 
 \phi_t ( d{y} ) 
 = 
 \frac{\partial \hat{C}}{\partial x} ( t , \hat{X}_t ) 
 \delta_U ( d{y}) 
 + 
 \left( 
 \hat{C} ( t , \hat{X}_t ) 
 - 
 \hat{X}_t \frac{\partial \hat{C}}{\partial x} ( t , \hat{X}_t ) 
 \right) 
 \delta_{T} (d{y}) 
. 
\end{equation} 
 Furthermore, when $(\hat{X}_t)_{t\in \real_+}$ is a geometric 
 Brownian motion given by \eqref{utfm} under $\hat{\P}$, 
 the bond call option with payoff 
$$ 
 (P_T( \mu)-\kappa P_T ( \nu ) )^+ 
 = 
 (P_T(U)-\kappa )^+ 
$$ 
 is priced as 
$$ 
 \E\left[ 
 e^{-\int_t^T r_s ds} 
 (P_T(U) - \kappa )^+ 
 \Big| 
 {\cal F}_t 
 \right] 
 = 
 P_t(U) 
 \Phi^0_+ ( t , \kappa , \hat{X}_t ) 
 - 
 \kappa 
 P_t(T) 
 \Phi^0_- ( t , \kappa , \hat{X}_t ) 
, 
$$ 
 and the corresponding hedging strategy is therefore given by 
\begin{equation} 
\label{cnc} 
 \phi_t ( d{y} ) 
 = 
 \Phi^0_+ ( t , \kappa , \hat{X}_t ) 
 \delta_U ( d{y}) 
 - \kappa 
 \Phi^0_- ( t , \kappa , \hat{X}_t ) 
 \delta_{T} (d{y}) 
, 
\end{equation} 
 from \eqref{c0a}. 
 When the dynamics of $(P_t)_{t\in \real_+}$ is given by 
 \eqref{cfp12} 
 where $\zeta_t (y)$ is deterministic, 
 $\hat{\sigma} (t)$ is given from 
 \eqref{asfvg} and 
 Lemma~\ref{l} 
 as 
$$ 
 \hat{\sigma} (t) 
 = 
 \zeta_t(U) 
 - 
 \zeta_t(T) 
, 
 \qquad 
 0 \leq t \leq T \leq U, 
$$ 
 and we check that \eqref{cnc} coincides with the result 
 \eqref{cnc0} obtained in Section~\ref{6}, cf. also 
 page 207 of \cite{privault-teng2}. 
\subsubsection*{Caplets} 
 Here we take $T<S$, 
 $X_t = P_t(\mu) = P_t(T)$, 
 $M_t = P_t(\nu) = P_t(S)$, 
 with 
 $$ 
 \mu (d{y}) = \delta_T (d{y}) 
 \quad 
 \mbox{ and } 
 \quad 
 \nu ( d{y} ) = \delta_S (d{y}) 
, 
$$ 
 and we consider the caplet with payoff \eqref{stl} 
 on the LIBOR rate \eqref{lts}, i.e. 
$$ 
 \xi 
 = 
 (S-T)(L(T,T,S)-\kappa )^{+} 
 = 
 ( \hat{X}_T - ( 1+\kappa (S-T) ) )^{+}. 
$$
 Assuming that $\hat{X}_t = P_t(T)/P_t(S)$ 
 is a (driftless) geometric Brownian motion under 
 $\hat{\P}$ with $\hat{\sigma} (t)$ 
 a deterministic function, this caplet is priced as in \eqref{bsf} as 
\begin{eqnarray*} 
\lefteqn{ 
 (S-T) 
 \E \left[ e^{-\int_t^S r_s ds} 
 (L(T,T,S)-\kappa )^{+} 
 \Big| {\cal F}_t 
 \right] 
} 
\\ 
 & = & 
 M_t 
 \hat{\E} \left[ 
 ( \hat{X}_T - ( 1+\kappa (S-T) ) )^{+} 
 \Big| {\cal F}_t 
 \right] 
\\ 
 & = & 
 P_t(T) 
 \Phi^0_{+}(t, 1+\kappa (S-T) , \hat{X}_t) 
 - 
 ( 1+\kappa (S-T) ) \Phi^0 _{-}(t, 1+\kappa (S-T) , 
 \hat{X}_t) P_t(S) 
, 
\end{eqnarray*} 
 since $P_S(\nu )=1$, and 
 the corresponding hedging strategy is given 
 as in \eqref{c0a} by 
\begin{equation} 
\label{dklddds} 
 \phi_t(dy)=\Phi^0_{+}(t, 1+\kappa (S-T) , 
 \hat{X}_t)\delta_T(dy)- ( 1+\kappa (S-T) ) 
 \Phi^0_{-}(t, 1+\kappa (S-T) , 
 \hat{X}_t)\delta _S (dy). 
\end{equation} 
 When the dynamics of $(P_t)_{t\in \real_+}$ is given by 
 \eqref{cfp12}, 
 where $\zeta_t (y)$ in \eqref{cfp12} is deterministic, 
 Lemma~\ref{l} shows that $ \hat{\sigma} (t)$ in \eqref{utfm} 
 can be taken as 
$$ 
 \hat{\sigma} (t) 
 = 
 \zeta_t(T) 
 - 
 \zeta_t(S) 
, 
 \qquad 
 0 \leq t \leq T \leq S, 
$$ 
 and in this case \eqref{dklddds} 
 coincides with Relation~\eqref{dklddds.1} above. 
\\ 
 
 Hedging strategies for caps are easily computed by 
 summation of hedging strategies for caplets. 
\subsubsection*{Swaptions on LIBOR rates} 
 Consider a tenor structure $\{ T \leq T_i, \ldots , T_j \}$ 
 and the swaption on the LIBOR rate with payoff 
\begin{equation} 
\label{qw} 
 \xi 
 = 
 P_T(\nu) 
 \hat{g} \left( 
 \frac{P_T(T_i) 
 - 
 P_T(T_j)}{ 
 P_T(\nu) 
 } 
 \right) 
, 
\end{equation} 
 where 
$$ 
 \hat{X}_t = 
 \frac{P_t(\mu)}{P_t(\nu) } 
 = 
 \frac{P_t(T_i) - P_t(T_j)}{P_t(\nu) }, 
 \qquad 
 0 \leq t \leq T, 
$$ 
 is the swap rate, which is a martingale under $\hat{\P}$, 
 in which case we have 
$$ 
 \mu (d{y}) 
 = 
 \delta_{T_i} (d{y}) - \delta_{T_j} (d{y}) 
 \quad 
 \mbox{and} 
 \quad 
 \nu (d{y}) =  
 \sum_{k=i}^{j-1} 
 \tau_k 
 \delta_{T_{k+1}} (d{y}) 
$$ 
 and 
$$ 
 M_t = P_t(\nu)  = 
 \sum_{k=i}^{j-1} 
 \tau_k 
 P_t(T_{k+1}) 
$$ 
 is the annuity numeraire. 
\\ 
 
 When $(\hat{X}_t)_{t\in \real_+}$ is Markov as in \eqref{mp}, 
 the self-financing hedging strategy of the 
 swaption with payoff \eqref{qw} is given 
 by \eqref{cf1} as 
\begin{eqnarray} 
\nonumber 
 \phi_t ( d{y} ) 
 & = & 
 \frac{\partial \hat{C}}{\partial x} ( t , \hat{X}_t ) 
 \delta_{T_i} ( d{y}) 
 + 
 \left( 
 \hat{C} ( t , \hat{X}_t ) 
 - 
 \hat{X}_t \frac{\partial \hat{C}}{\partial x} ( t , \hat{X}_t ) 
 \right) 
 \sum_{k=i+1}^{j-1} 
 \tau_{k-1} 
 \delta_{T_k} (d{y}) 
\\ 
\nonumber 
 & & 
 + 
 \left( 
 \tau_{j-1} 
 \hat{C} ( t , \hat{X}_t ) 
 - 
 ( 1 + \tau_{j-1} \hat{X}_t ) 
 \frac{\partial \hat{C}}{\partial x} ( t , \hat{X}_t ) 
 \right) 
 \delta_{T_j} (d{y}) 
, 
\end{eqnarray} 
 $0\leq t \leq T$. 
\\ 
 
 Finally we assume that the swap rate 
$$ 
 \hat{X}_t : = 
 \frac{P_t(T_i) - P_t(T_j)}{
 \sum_{k=i}^{j-1} 
 \tau_k 
 P_t(T_{k+1}) 
}, 
 \qquad 
 0 \leq t \leq T, 
$$ 
 is modeled according to 
 a driftless geometric Brownian motion under the forward swap 
 measure $\hat{\P}$ determined by 
 $M_t : =  \displaystyle \sum_{k=i}^{j-1} \tau_k P_t(T_{k+1})$, 
 $t\in \real_+$, with $(\hat{\sigma} (t) )_{t\in [0,T]}$ 
 a deterministic function. 
 In this case the swaption with payoff 
$$ 
 ( 
 P_T ( \mu ) 
 - 
 \kappa 
 P_T ( \nu ) 
 )^+ 
 = 
 ( 
 P_T(T_i) - P_T(T_j) 
 - 
 \kappa 
 P_T(\nu) 
 )^+, 
$$ 
 priced from \eqref{bsf} as 
\begin{eqnarray*} 
\lefteqn{ 
 \E\left[ 
 e^{-\int_t^T r_s ds} 
 ( 
 P_T(T_i) - P_T(T_j) 
 - 
 \kappa 
 P_T(\nu) 
 )^+ 
 \Big| 
 {\cal F}_t 
 \right] 
} 
\\ 
 & = & 
 ( 
 P_t(T_i) - P_t(T_j) 
 ) 
 \Phi^0_+ ( t , \kappa , \hat{X}_t ) 
 - 
 \kappa 
 P_t(\nu)  
 \Phi^0_- ( t , \kappa , \hat{X}_t ) 
\end{eqnarray*} 
 has the self-financing hedging strategy 
\begin{eqnarray} 
\nonumber 
 \phi_t ( d{y} ) 
 & = & 
 \Phi^0_+ ( t , \kappa , \hat{X}_t ) 
 \delta_{T_i} ( d{y}) 
 - 
 ( 
 \Phi^0_+ ( t , \kappa , \hat{X}_t ) 
 + 
 \kappa 
 \tau_{j-1} 
 \Phi^0_- ( t , \kappa , \hat{X}_t ) 
 ) 
 \delta_{T_j} (d{y}) 
\label{sw} 
\\ 
 & & 
 - \kappa 
 \Phi^0_- ( t , \kappa , \hat{X}_t ) 
 \sum_{k=i+1}^{j-1} 
 \tau_{k-1} 
 \delta_{T_k} (d{y}) 
, 
\end{eqnarray} 
 by \eqref{c0a}. 
 This recovers the self-financing hedging strategy 
\begin{equation} 
\label{eqp} 
 \Phi^0_+ ( t , \kappa , \hat{X}_t ) 
 \delta_{T_i} 
 - 
 \Phi^0_+ ( t , \kappa , \hat{X}_t ) 
 \delta_{T_j} 
 - 
 \kappa 
 \Phi^0_- ( t , \kappa , \hat{X}_t ) 
 \sum_{k=i}^{j-1} 
 \tau_k 
 \delta_{T_{k+1}} 
\end{equation} 
 of \cite{jamshidian2}, priced as 
$$ 
 \Phi^0_+ ( t , \kappa , \hat{X}_t ) 
 P_t(T_i) 
 - 
 \Phi^0_+ ( t , \kappa , \hat{X}_t ) 
 P_t ( T_j ) 
 - 
 \kappa 
 \Phi^0_- ( t , \kappa , \hat{X}_t ) 
 \sum_{k=i}^{j-1} 
 \tau_k 
 P_t( T_{k+1} ) 
$$ 
 The above hedging strategy \eqref{sw} shares the 
 same maturity dates as \eqref{dkldd111} above, although 
 it is stated in a different model. 

\footnotesize 

\def\cprime{$'$} \def\polhk#1{\setbox0=\hbox{#1}{\ooalign{\hidewidth
  \lower1.5ex\hbox{`}\hidewidth\crcr\unhbox0}}}
  \def\polhk#1{\setbox0=\hbox{#1}{\ooalign{\hidewidth
  \lower1.5ex\hbox{`}\hidewidth\crcr\unhbox0}}} \def\cprime{$'$}


\begin{thebibliography}{10}

\bibitem{carmonatehranchi}
R.~A. Carmona and M.~R. Tehranchi.
\newblock {\em Interest rate models: an infinite dimensional stochastic
  analysis perspective}.
\newblock Springer Finance. Springer-Verlag, Berlin, 2006.

\bibitem{corcuera}
J.M. Corcuera.
\newblock Completeness and hedging in a {L}\'evy bond market.
\newblock In A.~Kohatsu-Higa, N.~Privault, and S.J. Sheu, editors, {\em
  Stochastic Analysis with Financial Applications (Hong Kong, 2009)}, volume~65
  of {\em Progress in Probability}, pages 317--330. Birkh\"auser, 2011.

\bibitem{daprato}
G.~Da~Prato and J.~Zabczyk.
\newblock {\em Stochastic equations in infinite dimensions}.
\newblock Encyclopedia of Mathematics and its Applications. Cambridge
  University Press, Cambridge, 1992.

\bibitem{filipovicbk}
D.~Filipovi{\'c}.
\newblock {\em Consistency problems for {H}eath-{J}arrow-{M}orton interest rate
  models}, volume 1760 of {\em Lecture Notes in Mathematics}.
\newblock Springer-Verlag, Berlin, 2001.

\bibitem{geman}
H.~Geman, N.~El Karoui, and J.-C. Rochet.
\newblock Changes of num\'eraire, changes of probability measure and option
  pricing.
\newblock {\em J. Appl. Probab.}, 32(2):443--458, 1995.

\bibitem{cfhuang}
C.-F. Huang.
\newblock Information structures and viable price systems.
\newblock {\em Journal of Mathematical Economics}, 14:215--240, 1985.

\bibitem{jamshidian2}
F.~Jamshidian.
\newblock Sorting out swaptions.
\newblock {\em Risk}, 9(3):59--60, 1996.

\bibitem{jamshidian4}
F.~Jamshidian.
\newblock Numeraire invariance and application to option pricing and hedging.
\newblock MPRA Paper No. 7167, 2008.

\bibitem{ko:gc}
I.~Karatzas and D.L. Ocone.
\newblock A generalized {C}lark representation formula with application to
  optimal portfolios.
\newblock {\em Stochastics and Stochastics Reports}, 34:187--220, 1991.

\bibitem{privault-teng2}
N.~Privault and T.-R. Teng.
\newblock Risk-neutral hedging in bond markets.
\newblock {\em Risk and Decision Analysis}, 3:201--209, 2012.

\bibitem{protterspa}
P.~Protter.
\newblock A partial introduction to financial asset pricing theory.
\newblock {\em Stochastic Process. Appl.}, 91(2):169--203, 2001.

\bibitem{schoenmakersbk}
J.~Schoenmakers.
\newblock {\em Robust {LIBOR} modelling and pricing of derivative products}.
\newblock Chapman \& Hall/CRC Financial Mathematics Series. Chapman \&
  Hall/CRC, Boca Raton, FL, 2005.

\end{thebibliography}
\end{document}